\documentclass[a4paper,notitlepage,10pt]{article}

\usepackage[utf8]{inputenc}
\usepackage[LGR,T1]{fontenc}        
\usepackage[british]{babel}
\usepackage[iso,british]{isodate}           

\usepackage[top=2.5cm,bottom=2.5cm,left=2cm,right=2cm]{geometry} 
\usepackage[compact]{titlesec}	    
\usepackage[bottom]{footmisc}       

\usepackage[babel=true,final]{microtype}	
\usepackage{parskip}                        
\usepackage{mlmodern}
\usepackage{textgreek}
\usepackage{hyphenat}

\usepackage{graphicx}               
\usepackage{float}                  
\usepackage{caption}
\usepackage{subcaption}
\usepackage{amsmath}
\usepackage{abstract}
\usepackage{authblk}
\usepackage{enumitem}
\usepackage{siunitx}
\usepackage{xpunctuate}

\newcommand{\ie}        {\emph{i.e.}\xspace}
\newcommand{\eg}        {\emph{e.g.}\xspace}
\newcommand{\etal}      {\emph{et al}\xperiod}

\newcommand{\CA}         {\theta}
\newcommand{\DP}      [1]{\Delta P_{\rm #1}}
\newcommand{\DPbrush}    {\Delta P^{\rm brush}_{\graft,\eps{pl}}}

\newcommand{\DPgraft}    {\Delta P^{\rm graft}}

\newcommand{\eps}     [1]{\epsilon_{\rm #1}}
\newcommand{\graft}      {\rho_g}
\newcommand{\graftN}     {N_g}
\newcommand{\Hght}       {H}
\newcommand{\IntEne}  [1]{\gamma_{\rm #1}}
\newcommand{\IntTns}  [1]{\Upsilon_{\rm #1}}

\newcommand{\Ptnsr}      {\mathbf{P}}
\newcommand{\rij}        {r_{ij}}
\newcommand{\Young}      {\theta_{\rm y}}
\newcommand{\za}         {z_{\rm a}}
\newcommand{\zb}         {z_{\rm b}}

\usepackage[sorting=none,sortcites,giveninits,maxnames=5,backend=biber,urldate=iso,date=year,citestyle=numeric-comp]{biblatex} 
\usepackage{csquotes}				

\usepackage[dvipsnames]{xcolor}		
\usepackage[a-3u]{pdfx}

\newcommand{\SKIP} [1]{}

\definecolor{CoolBlack}{rgb}{0.0, 0.18, 0.39}
\hypersetup{
	hidelinks,
	colorlinks,
	linkcolor  = CoolBlack,
	citecolor  = CoolBlack,
	urlcolor   = OliveGreen
}

\newcommand*\diff{\mathop{}\!\mathrm{d}}

\title{Pressure anisotropy in polymer brushes and its effects on wetting}
\author[1, *]{Lars B. Veldscholte}
\author[2]{Jacco H. Snoeijer}
\author[3]{Wouter K. den Otter}
\author[1]{Sissi de Beer}
\affil[1]{Functional Polymer Surfaces, Department of Molecules and Materials, MESA+ Institute, University of Twente}
\affil[2]{Physics of Fluids, MESA+ Institute, University of Twente}
\affil[3]{Multiscale Mechanics, Department of Fluid and Thermal Engineering, MESA+ Institute, University of Twente}
\affil[*]{Corresponding author: l.b.veldscholte@utwente.nl, P.O. Box 217, 7500 AE Enschede, the Netherlands}
\date{\today}

\begin{document}

\twocolumn[\maketitle\begin{abstract}
    \vspace{2em}
    Polymer brushes, coatings consisting of densely grafted macromolecules, experience an intrinsic lateral compressive pressure, originating from chain elasticity and excluded volume interactions.
    This lateral pressure complicates a proper definition of the interface and, thereby, the determination and interpretation of the interfacial tension and its relation to the wetting behavior of brushes.
    Here, we study the link between grafting-induced compressive lateral pressure in polymer brushes, interfacial tension, and brush wettability using coarse-grained molecular dynamics simulations. We focus on grafting densities and polymer-liquid affinities such that the polymer and liquid do not tend to mix.
    For these systems, a central result is that the liquid contact angle is independent of the grafting density, which implies that the grafting-induced lateral compressive pressure in the brush does not influence its wettability.
    Although the definition of brush interfacial tensions is complicated by the grafting-induced pressure, the difference in interfacial tension between wet and dry brushes is perfectly well-defined. We confirm explicitly from Young's law that this difference offers an accurate description of the brush wettability. 
    We then explore a method to isolate the grafting-induced contribution to the lateral pressure, assuming the interfacial tension to be independent of grafting density. This scenario indeed allows to disentangle interfacial and grafting effects for a broad range of parameters, except close to the mixing point. We separately discuss the latter case in the light of autophobic dewetting.
	\vspace{2em}
\end{abstract}]


\section{Introduction}
Polymer brushes are coatings consisting of macromolecules that are end-grafted to a substrate at sufficiently high grafting densities that they are forced to stretch away. They then form a so-called `brush' structure, with the polymer chains extending upwards from the substrate.
These systems show considerably different behaviors compared to bulk polymers or non-grafted films. Coatings consisting of polymer brushes have several applications in stabilization of colloids~\cite{Pincus1991}, sensors~\cite{Merlitz2009,Klushin2014}, separation membranes~\cite{Keating2016,Pizzoccaro2019,Bilchak2020,Durmaz2021}, low\hyp{}friction~\cite{deBeer2014,Bhairamadgi2014} and anti\hyp{}fouling coatings~\cite{Benetti2019}, and stimulus\hyp{}responsive materials~\cite{chen_50th_2017} such as smart adhesives~\cite{Yu2021}. In several of these applications, polymer brushes are applied in air~\cite{ritsema_van_eck_fundamentals_2022}, as opposed to immersed in liquid. In these cases, understanding of wetting behavior of polymer brushes is imperative, especially since brushes are known to display counter-intuitive wetting effects~\cite{Cohen2006,Mensink2019,leonforte_statics_2011}.

\begin{figure}[tb]
  \centering
  \includegraphics[width=0.75\linewidth]{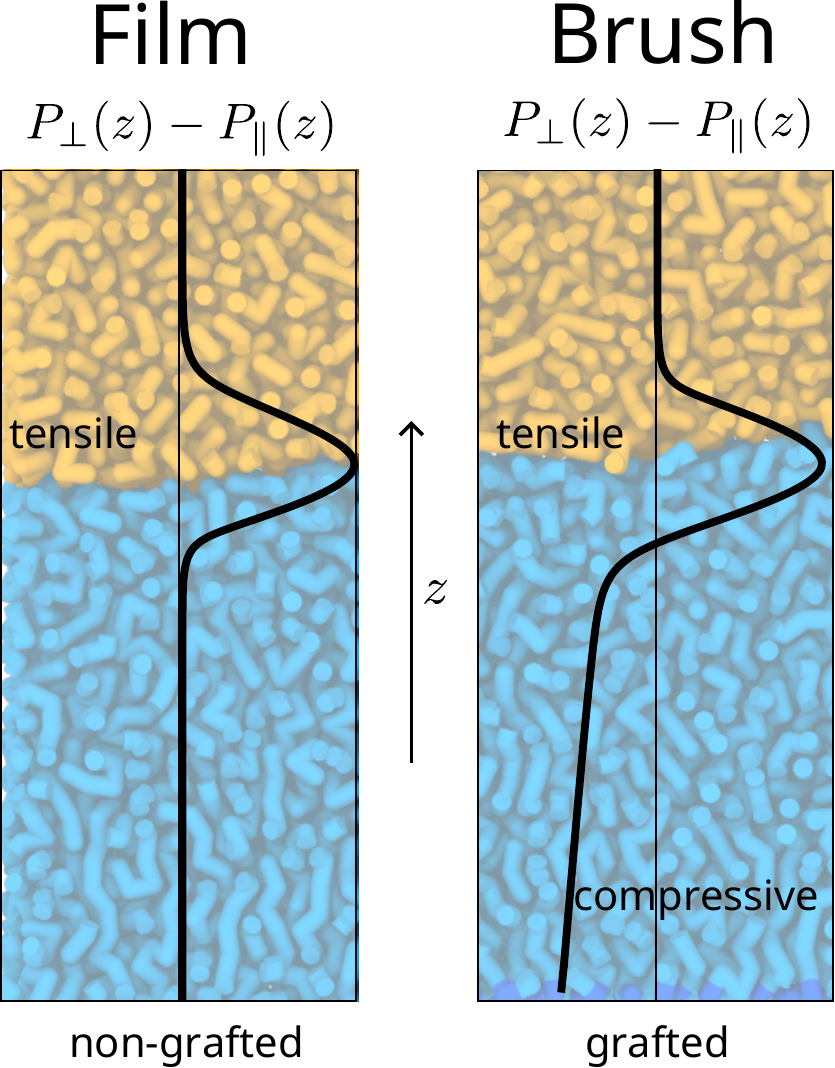}
  \caption{Illustrations of pressure anisotropy profiles (left) for a liquid on top of a non-grafted film, and (right) for a liquid on top of a brush. In both cases a tensile lateral pressure localized near the interface, whose integral is naturally associated with the surface tension, arises. The pressure anisotropy profile for a liquid on top of a brush shows an additional grafting-induced compressive pressure within the brush.}
  \label{fig:ga}
\end{figure}

A polymer brush's conformation is the result of a competition between the polymer chains' entropic elasticity, which tends to contract the brush, and excluded volume interactions between the chains, which are repulsive and thereby extend the brush.
This competition also produces an inherent lateral compressive pressure in polymer brushes: the pressure inside the brush is anisotropic, with an effective compression parallel to the grafting plane.
As sketched in \autoref{fig:ga}, this pressure is largest at the grafting plane and smoothly reaches zero at the brush surface, as shown by both self\hyp{}consistent field theory \cite{manav_mechanics_2018} and molecular dynamics simulations \cite{dimitrov_polymer_2007, manav_stress_2019}.
This grafting\hyp{}induced lateral pressure can even bend a flexible substrate \cite{manav_mechanics_2018, manav_stress_2021}.
No such lateral pressure exists in non-grafted polymer films.

\begin{figure*}[tb]
    \centering
    \includegraphics[height=37mm]{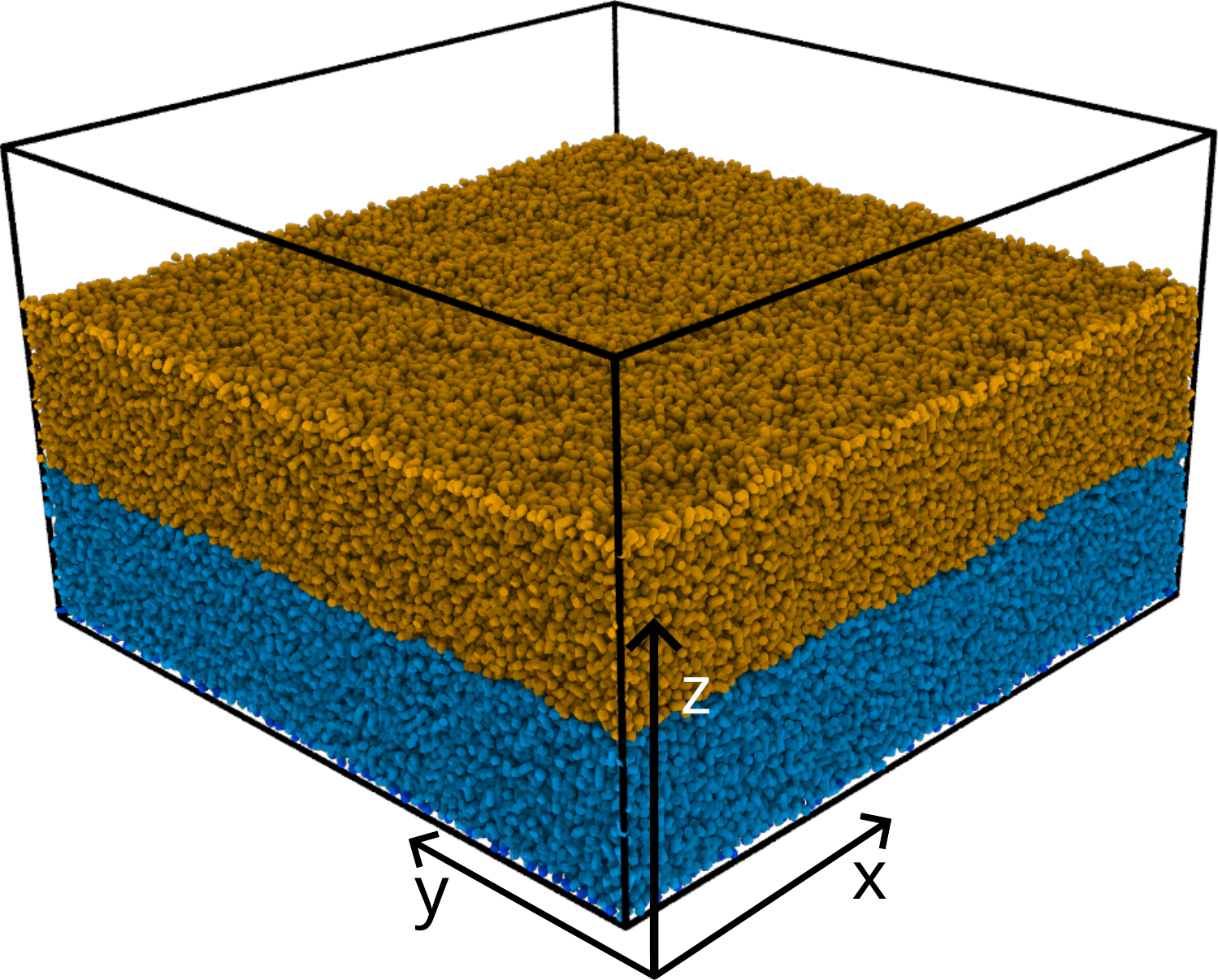}
    \hfill
    \includegraphics[height=37mm]{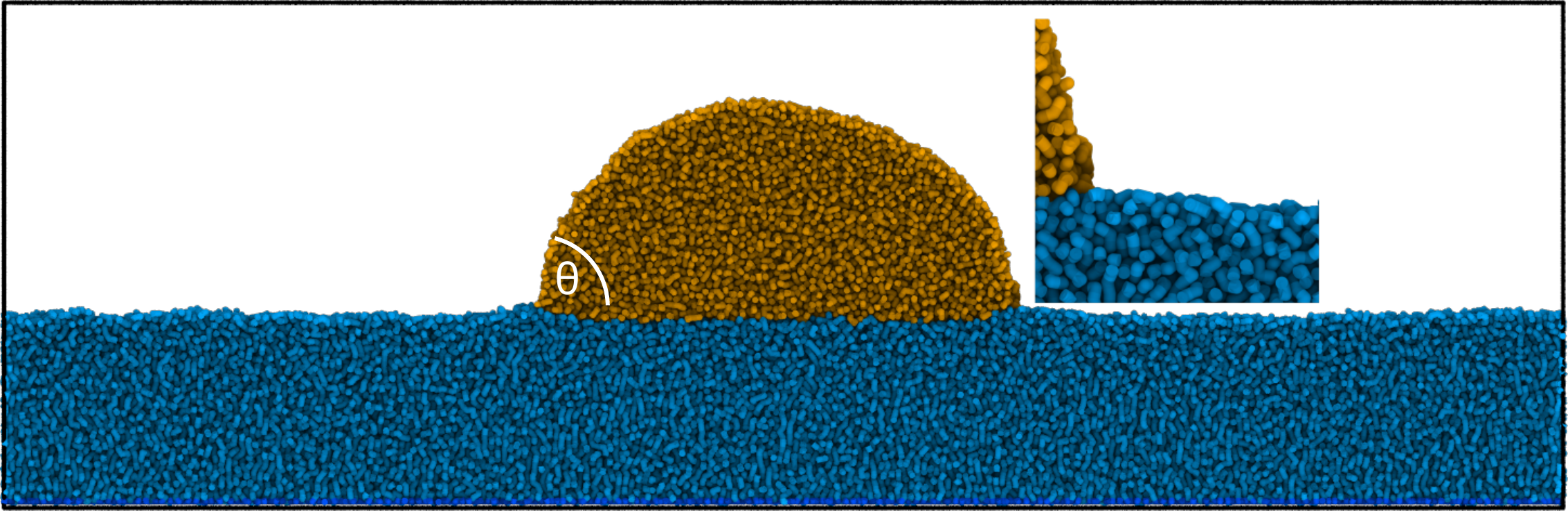}
    \caption{
        Snapshots of the two types of simulation setups. (Left) A liquid layer on top of a brush or film in a box with a square ground plane. (Right) Cross-section of a cylindrical droplet on a brush or film in a rectangular quasi-2D setup. The inset zooms in on the contact line, highlighting the absence of a wetting ridge. The smooth wall supporting the brush or film from below is not shown.}
    \label{fig:setups}
\end{figure*}

Polymer brushes are able to ad- and absorb solvent, just like non-grafted polymer films. Whereas the latter will dissolve in a good solvent, the grafted polymers of a brush can merely swell.
An interesting situation arises when the solvent conditions are less favorable, such that an interface will form. An interfacial tension will emerge, which manifests itself as a lateral tensile pressure localized near the interface, as illustrated in \autoref{fig:ga}. To make this more explicit, we define the pressure anisotropy as
\begin{alignat}{3}
    &\DP{}(z) &&= P_\perp(z) &&- \quad\quad\quad P_\parallel(z) \nonumber \\
    & &&= P_{zz}(z) && - \frac12 \Big[P_{xx}(z) + P_{yy}(z)\Big],
\end{alignat}
where $P_{xx}$, $P_{yy}$ and $P_{zz}$ denote the three diagonal elements of the Cartesian pressure tensor $\Ptnsr$, with $P_\perp$ and $P_\parallel$ the pressures perpendicular and parallel, respectively, to the flat substrate. The definitions of the Cartesian directions are given in \autoref{fig:setups}, in a 3D snapshot of a Molecular Dynamics (MD) simulation.
In this set-up, the planar symmetry of the system means that the pressure parallel to the interface is a function of $z$ only, $P_\parallel(z) = P_{xx}(z) = P_{yy}(z)$, while mechanical equilibrium dictates a constant normal pressure equal to the isotropic bulk pressure in the fluid, $P_\perp = P_{zz} = P_{\rm fluid}$ \cite{rowlinson_molecular_2002, varnik_molecular_2000}.
While it is difficult to measure the pressure distribution in a brush experimentally, it can be extracted from molecular dynamics (MD) simulations \cite{Goetz_1998, ikeshoji_molecular-level_2003} with relative ease.
The interfacial tension is then obtained as the integral of $\Delta P(z)$ over the interface \cite{frenkel_understanding_2001, rowlinson_molecular_2002, dimitrov_polymer_2007, andreotti_statics_2020},
\begin{align}
    \IntTns{}
  =
    \int_{\za}^{\zb}
      \left[ P_\perp(z) - P_\parallel(z) \right]
    \diff z,
  \label{eq:interfacial_tension}
\end{align}
where the integration boundaries $\za$ and $\zb$ are located in the bulk phases on either side of the interface.

The above equation for the interfacial tension offers a clear physical interpretation for the interface between two immiscible liquids, where the pressure anisotropy is non-zero only in a small region near the interface. The same applies for the interface between a polymer melt and a liquid. For a brush-liquid interface, however, it is less clear what exactly constitutes the interfacial region. This can be seen from the sketched pressure anisotropy profiles in \autoref{fig:ga}: the pressure anisotropy is negative (compressive) in the brush bulk, and grows more negative with increasing distance from the interface \cite{dimitrov_polymer_2007}. Therefore, it is not straightforward to determine the actual ``interfacial excess'' that defines the thermodynamic property of the interface.
Instead, we are left with the unpleasant situation where the value of the interfacial tension varies with the choice of the integration boundary in the brush, $\za$.

The compressive pressure anisotropy within the brush is due to grafting, not due to interfacial interactions, and hence it would be inappropriate to treat the full integral as a traditional liquid-liquid interfacial tension. Including (part of) the brush bulk can even yield a negative interfacial tension if the grafting-induced compressive pressure is high enough compared to the peak in the interfacial region. This result has led to various interpretations.

For example, Dimitrov \etal~\cite{dimitrov_polymer_2007} include the entire brush in the surface tension calculation and conclude that this surface tension is fundamentally different from a conventional liquid-liquid interfacial tension: it is not limited to positive values, and a change in sign is not accompanied by a phase transition.
Moreover, they propose that the pressure anisotropy profile can be separated in a `brush' and an `interfacial' part by splitting at the height $z_0$ where the anisotropy vanishes, $\Delta P(z_0) = 0$.
Badr \etal~\cite{badr_cloaking_2022} ``choose to evaluate only the integral over the peaks at the interfaces''.
Similarly, Milchev \etal~\cite{milchev_concave_2023} noted negative interfacial tensions that grow in absolute value with increasing grafting density in concave polymer brushes.
Léonforte and Müller~\cite{leonforte_statics_2011} extracted interfacial tensions from the thermal fluctuation spectrum of the brush-vapor interface, reporting good agreement with the surface tensions of their non-grafted counterparts.
Whereas integrating only a selected part of the pressure anisotropy curve yields promising results, it is not evident whether this approach includes all interfacial contributions, nor whether it includes interfacial contributions only.

In this paper we will explore the link between the negative pressure anisotropy in brushes and the interfacial tensions, and how this affects the wettability of the brush.
We will use coarse-grained molecular dynamics simulations, focusing on the two types of configurations, illustrated in \autoref{fig:setups}: a layer of liquid deposited on top of the brush, resulting in a planar interface, and an infinitely long cylindrical droplet on top of the brush.
The former simulation yields interfacial tensions, and the latter probes the wettability by means of the contact angle of the droplet. Finally, we will compare the results against similar configurations of non-grafted polymer films, to explore whether the pressure anisotropy due to grafting and due to the presence of the interface can be disentangled.

\section{Surface tensions, surface energies, and wetting}
In spite of difficulties associated with defining interfacial tensions of brushes, their observable wetting properties should be well defined, and are only a function of the difference between the brush-liquid and brush-vapor interfacial energies. This follows from Young's law, which describes the contact angle of a liquid droplet (L) resting on a solid substrate (S) in equilibrium with its vapor (V):
\begin{equation}
    \cos(\Young) = \frac{ \IntEne{SV} - \IntEne{SL} }{ \IntEne{LV} },
    \label{eq:Young}
\end{equation}
where the $\IntEne{}$ denote interfacial free energies per unit area and the two subscripts specify the interface.

Two possible complications may arise here. First, Young's law applies to rigid surfaces, which polymer brushes and polymer films are not.
The brushes simulated here, however, are sufficiently stiff that Young's law can be used to estimate contact angles \cite{andreotti_statics_2020}.
Second, for elastic substrates the interfacial free energy density $\gamma$ is not necessarily equal to the interfacial tension $\Upsilon$ as defined in \autoref{eq:interfacial_tension}. The two quantities are related by $\diff( \IntEne{} A ) = \IntTns{} \diff A$, where $A$ denotes the interfacial area.
For simple liquids, these quantities always take on the same value, \ie $\gamma = \Upsilon$, and one does not explicitly distinguish between the two concepts.
In general, however, a local stretching of the interface can change the interfacial composition and thus alter its free energy density.
For elastic substrates this phenomenon is known as the Shuttleworth effect \cite{shuttleworth_surface_1950, andreotti_soft_2016,andreotti_statics_2020,salez_stretching_2020,style_elastocapillarity_2017}, while a similar distinction between interfacial tension and energy arises for surfactants~\cite{thiele_equilibrium_2018}.
When stretching a brush laterally, a constant number of grafting points $\graftN$ becomes distributed over a larger area, and consequently the grafting density $\graft = \graftN / A$ decreases.
The change in interfacial free energy is evaluated as
\begin{align}
    \diff(\gamma A)
    &=
    \gamma \diff A + A \diff \gamma
    =
    \gamma \diff A
    +
    A \frac{\diff \gamma}{\diff \graft} \frac{ \diff \graft }{ \diff A} \diff A,
\end{align}
and it follows that the interfacial tension $\Upsilon$ is related to the interfacial free energy density $\gamma$ by
\begin{equation}\label{eq:shuttleworth}
    \Upsilon = \gamma - \graft \frac{\diff \gamma}{\diff \graft }.
\end{equation}
By integrating the pressure anisotropy as outlined in \autoref{eq:interfacial_tension}, one obtains the interfacial tension, which for brushes does not necessarily equal the interfacial energy.

Previous simulations by Léonforte \etal \cite{leonforte_molecular_2011} of droplets on brushes of intermediate grafting densities show the contact angle to be independent of the grafting density. This observation suggests that $\diff \gamma/\diff \rho_g$ must either vanish, or at least be identical for bush-liquid (SL) and brush-vapor (SV) interfaces. Under this convenient condition, one finds 
\begin{align}
   \begin{split}
        \IntEne{LV}
        \cos{\Young}
    &= 
        \IntEne{SV} - \IntEne{SL}
     =
        \IntTns{SV} - \IntTns{SL}
\\
    &=
        \int_{\za}^{\zb}
        \Big[ \DP{SV}(z) - \DP{SL}(z) \Big]
        \diff z.
    \label{eq:IntDPdiff}
    \end{split}
\end{align}
This direct relation between 
    contact angles and pressure-anisotropy profiles
  will be explored below.


\section{Model and methods}
\subsection{Model}

Molecular dynamics simulations were performed using the LAMMPS~\cite{thompson_lammps_2021} MD software.
The chemically-aspecific coarse-grained Kremer-Grest model \cite{kremer_dynamics_1990} is used to represent polymers as freely-jointed bead-spring chains. The non-bonded interaction between particles $i$ and $j$ is described by the Lennard-Jones potential,
\begin{equation}\label{eq:LJ}
    U^{\rm LJ}_{ij}(r_{ij})
  =
    4 \epsilon_{ij}
    \left[ \left(\frac{\sigma_{ij}}{\rij}\right)^{12}
         - \left(\frac{\sigma_{ij}}{\rij}\right)^6
    \right],
\end{equation}
where $\rij$ is the interparticle distance, $\epsilon_{ij}$ represents the depth of the energy well and all particle pairs share a zero-crossing distance $\sigma_{ij} = \sigma$.
The potential is truncated at $r_{\rm c} = 2.5\,\sigma$,
\begin{align}
    U^{\rm LJ, cut}_{ij}(\rij)
 &=
    \begin{cases}
      U^{\rm LJ}_{ij}(\rij) - U^{\rm LJ}_{ij}(r_{\rm c})
    &
      \text{for } \rij \leq r_{\rm c}
    \\
      0
    &
      \text{for } \rij > r_{\rm c}.
    \end{cases}.
\end{align}
Bonded interactions between consecutive beads along a polymer chain are described by the purely attractive Finitely Extensible Non-linear Elastic (FENE) potential
\begin{align}
    U_{\rm FENE}(\rij)
 &=
  - \frac12 K R_0^2 \ln\left[1-\left(\frac{\rij}{R_0}\right)^2\right],
 \label{eq:fene}
\end{align}
combined with the purely repulsive Weeks-Chandler-Anderson (WCA) potential
\begin{align}
    U^{\rm WCA}_{ij}(\rij)
 &=
    \begin{cases}
    	U^{\rm LJ}_{ij}(\rij) + \epsilon_{ij}
    &
      \text{for } \rij \leq 2^{1/6}\,\sigma
    \\
      0
    &
      \text{for } \rij > 2^{1/6}\,\sigma.
    \end{cases}
  \label{eq:wca}
\end{align}
The spring constant $K = 30\,\epsilon/\sigma^{2}$ and the maximum bond length $R_0 = 1.5\,\sigma$ of the FENE potential are taken from the Kremer-Grest model~\cite{kremer_dynamics_1990}.
Throughout this work, we will use reduced Lennard-Jones units \cite{frenkel_understanding_2001} with $\epsilon$ and $\sigma$ serving as the unit of energy and length, respectively.
Time is expressed in units of $\tau = \sigma\sqrt{m/\epsilon}$, where $m$ is the particle mass, and temperatures are given in $\epsilon / k_{\rm B}$, with $ k_{\rm B}$ the Boltzmann constant.

In addition to the polymer, a liquid consisting of Kremer-Grest tetramers is introduced into the system. We use tetramers instead of monomers to depress the vapor pressure, thereby reducing the vapor density to virtually zero. Hence, the vapor looks empty in the snapshots of the system presented here.
The liquid beads are of the same size and mass as the polymer beads, and also interact with each other and with the polymer by a Lennard-Jones potential. The liquid (l) and polymer (p) self-interaction energies are set at $\eps{ll} = \eps{pp} = 1.5\,\epsilon$. They were given identical values for simplicity; the value of $1.5\,\epsilon$ was chosen to make the polymer sufficiently stiff to suppress the formation of a wetting ridge at the contact line (see the inset to \autoref{fig:setups}). To control the affinity between polymer and solvent, and thereby the wettability, the polymer-solvent interaction energy $\eps{pl}$ is varied between $0.5\,\epsilon$ and $1.5\,\epsilon$, resulting in contact angles of \SIrange{0}{130}{\degree}. Note that these Lennard-Jones interactions effectively capture all kinds of interactions between polymer beads, rather than merely the van der Waals interaction between two atoms, so combining rules are not applicable here and the $\epsilon_{ij}$ parameters can be varied independently of each other \cite{israelachvili_intermolecular_2015}.

A system consisting of a rectangular box with periodic boundary conditions along $x$ and $y$ is set up. Two different box sizes where used: a 3D system of Cartesian dimensions $100\,\sigma \times 100\,\sigma \times 60\,\sigma$ for the slab simulations, and a quasi-2D geometry of $250\,\sigma \times 10\,\sigma \times 80\,\sigma$ for the droplet simulations.
In the latter system, the droplet is periodically continued in the $y$ direction to create a cylindrical droplet and thereby eliminate line tension contributions to the contact angle, while the periodic repeat distance is still sufficiently short to suppress Plateau-Rayleigh instabilities \cite{weijs_origin_2011,law_line_2017,leonforte_statics_2011}.
\autoref{fig:setups} shows snapshots of both simulation setups, rendered using Ovito~\cite{stukowski_visualization_2009}.
The brushes and polymer films rest on a flat structureless mathematical wall (w) exerting a Lennard-Jones 9-3 potential with $\eps{pw} = 1\,\epsilon$ and $\sigma_{\rm pw} = 1\,\sigma$, where the zero-crossing height defines $z=0$.
At the top of the box a repulsive harmonic potential with spring constant of $100\,\epsilon/\sigma^2$ prevents evaporated fluid molecules from escaping.
A monodisperse polymer brush is created by `grafting' polymer chains of $N=50$ beads to immobile grafting beads positioned randomly in the $z=0$ bottom plane of the box.
This chain length was chosen as a balance between computational cost and adequate reproduction of polymer brush behavior.
The density of grafting points varies between $0.2$ and $0.6\,\sigma^{-2}$, which ensures that all our systems are in the brush regime. Specifically, brushes will form `holes' or crystallize for densities below or above this range, respectively.

\subsection{Simulation procedure}

Data files comprising initial configurations of fully-stretched Kremer-Grest polymer brushes in rectangular boxes at various grafting densities are generated using a Python script, made available online \cite{l_b_veldscholte_2020_3944454}.

The systems were equilibrated by a short energy minimization using the conjugate gradient method (\texttt{min\_style cg}), followed by a dynamics run for $250\,\tau$ with a displacement limit  (\texttt{fix nve/limit}) of $1\,\sigma$ per timestep.
The equilibration is performed with a Langevin thermostat (\texttt{fix langevin}) with a temperature of $1.7\,\epsilon/k_{\rm B}$ and a damping parameter of $10\,\tau$.
This was followed by a longer dynamics run for $5000\,\tau$ without the limit and a less viscous Langevin thermostat (damping parameter of $100\,\tau$) that cooled the system down to a temperature of $0.85\,\epsilon / k_{\rm B}$.
This procedure was chosen to equilibrate the polymer brush system as efficiently as possible.
In these equilibration runs, a time step of $0.005\,\tau$ was used.

For the production runs, the rRESPA multi-time-scale integrator \cite{tuckerman_reversible_1992} is used with an outer time step of $0.010\,\tau$ and a twofold smaller inner time step of $0.005\,\tau$. This results in non-bonded pair interactions being computed every $0.010\,\tau$, but bonded interactions being computed twice as often. Equilibrium simulations sample the canonical ensemble by thermostatting to a temperature of $0.85\,\epsilon / k_{\rm B}$ using a Nosé-Hoover chain (\texttt{fix nvt}) with a damping parameter of $0.1\,\tau$. Production runs of the brushes were run for $10^4\,\tau$, storing density and pressure profiles over $z$ at regular intervals.
Subsequently, either slabs or cylindrical droplets of liquid were deposited on top of the brushes.
The simulations with slabs were run for $10^4\,\tau$, storing density and pressure profiles over $z$, while droplets were simulated for $10^5\,\tau$, storing 2D density profiles over $x$ and $z$.
A similar procedure was applied for the polymer films; the number of polymer chains in these films corresponded with a grafting density of $\graft = 0.4\,\sigma^{-2}$.
To ensure only data from systems in equilibrium are included in the analyses, the first half of each production run was discarded and only the latter half was analyzed.

\subsection{Data analysis}
Computing the pressure anisotropy profile $\DP{}(z)$ necessitates a definition of the local pressure tensor, which is not uniquely defined for inhomogeneous systems \cite{Schofield_1982,varnik_molecular_2000, hafskjold_microscopic_2002, rowlinson_molecular_2002, smith_importance_2022, ikeshoji_molecular-level_2003}.
The kinetic pressure contribution by particle $i$ is readily assigned to the bin along the $z$ direction holding the particle. For the virial contribution by the interaction between the $i-j$ particle pair we used the Irving-Kirkwood contour: the contribution is distributed over all bins from $i$ to $j$ in proportion to the fraction of the total distance from $i$ to $j$ traversed in each of these bins \cite{Goetz_1998, ikeshoji_molecular-level_2003}.
This distribution of the virial over the bins has the appealing feature of $P_{\perp}(z)$ being constant, in agreement with mechanical equilibrium along the $z$ direction \cite{varnik_molecular_2000, ikeshoji_molecular-level_2003}. We note that the integrals of the pressure profiles $P_{\perp}(z)$ and $P_{||}(z)$ over the entire height of the simulation box are independent of the chosen distribution.
Galteland \etal \cite{galteland_nanothermodynamic_2021} implemented this algorithm in LAMMPS (\texttt{compute stress/cartesian}) for particles interacting by non-bonded pair forces. We amended the routine to include the bond forces in our polymers; this code is merged in LAMMPS and available as of the 15 June 2023 release.

Contact angles are extracted from the 2D density profiles of droplets by circle fitting, in the spirit of \cite{weijs_origin_2011, mensink_role_2021, carrillo_adhesion_2010, badr_cloaking_2022}. Data processing, analysis and visualization were performed using Python, with NumPy \cite{harris2020array}, SciPy \cite{2020SciPy-NMeth} and Matplotlib \cite{Hunter:2007}. The Python code, \eg to parse LAMMPS \texttt{ave/chunk} output or to extract contact angles from droplets, is made available online \cite{lars_veldscholte_2022_6341254}.

\section{Results and discussion}
First, we will look at the contact angle of a droplet on a polymer brush as a function of polymer-liquid affinity, for several values of the brush grafting density. Next, we address the pressure anisotropy profiles of brushes and films, and examine the influence of the integration limit on the interfacial tensions and resulting Young's contact angle. We then compare the measured contact angles with the Young's contact angles. Finally, we explore a route to disentangle the interfacial and bulk contributions to the pressure anisotropy.

\newpage
\subsection{Contact angles of droplets on brush\-es of varying grafting density}
\begin{figure}[tb]
    \centering
    \includegraphics{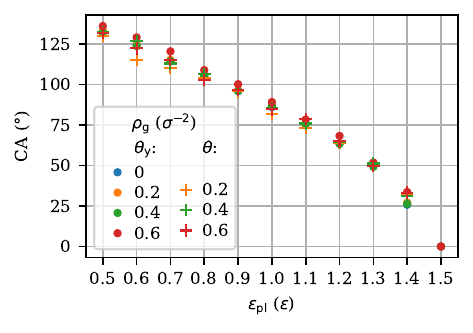}
    \caption{
        Contact angles as a function of polymer-liquid interaction strength $\eps{pl}$, for polymer brushes at the grafting densities $\graft$ indicated in the legend and for an non-grafted film denoted by $\graft = 0$. Pluses represent contact angles $\CA$ measured from simulations of droplets, dots represent contact angles $\Young$ calculated from the interfacial tensions using Young's law.
        The droplets at $\eps{pl} = 1.5\,\epsilon$ are very wide and shallow, rather than the nearly-cylindrical shape observed at smaller $\eps{pl}$, which suggests that they are perfectly wetting, $\CA = 0^\circ$.
        Contact angles for droplets on non-grafted polymer films are not included because they deviate from Young's law by showing Neumann wetting: the droplet curves the interface underneath and near the droplet.
    }
    \label{fig:CAvsEPS}
\end{figure}

Now, we can test the validity of \autoref{eq:IntDPdiff}, \ie the independence of contact angles of droplets from the brush's grafting density. The contact angles $\CA$ measured from the simulations of droplets on brushes over a range of polymer-liquid interaction strengths $\eps{pl}$ are represented in \autoref{fig:CAvsEPS} by pluses. The markers at different grafting densities $\graft$ essentially overlap, indicating that the contact angle of a droplet on the brush is independent of its grafting density for these liquid-brush combinations, in line with a previous study by Léonforte \etal \cite{leonforte_molecular_2011}. That is, the difference in surface energies $\IntEne{SV} - \IntEne{SL}$ is independent of $\graft$. This implies that the polymer-liquid $\gamma_{\rm SL}$ and polymer-vacuum $\gamma_{\rm SV}$ surface energies either are not dependent on the grafting density, in which case there is no Shuttleworth effect, or they have exactly the same grafting density dependence, in which case the Shuttleworth effect is called symmetric \cite{henkel_soft_2022}.
In view of \autoref{eq:shuttleworth}, this conveniently implies that in the evaluation of differences between solid-liquid and solid-vapor interfaces, such as in Young's law, we do not have to distinguish between surface tension $\IntTns{}$ and surface free energy density $\IntEne{}$, which simplifies the discussion.

\begin{figure*}[tb]
    \centering
    \includegraphics{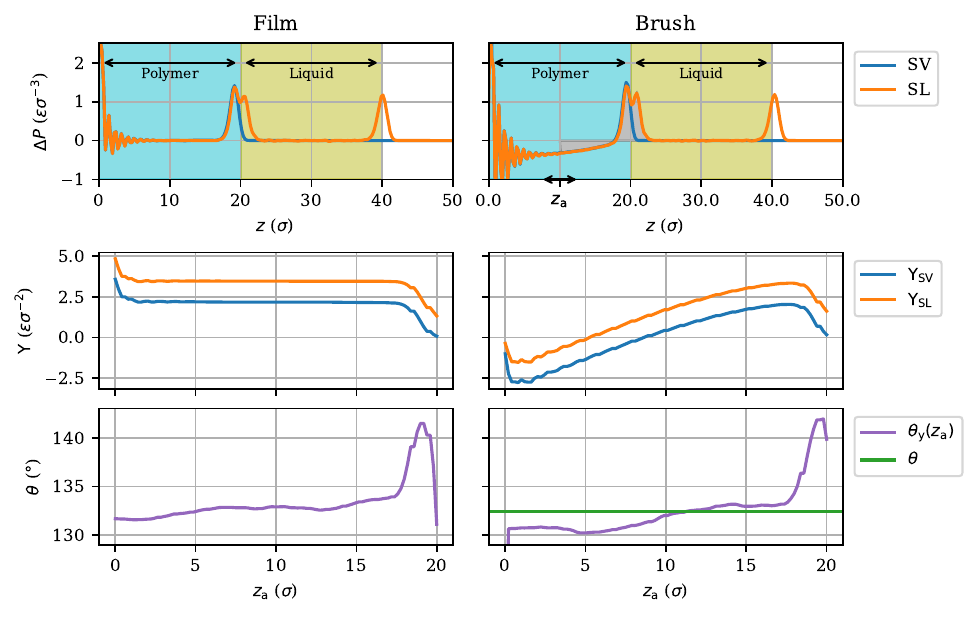}
    \caption{
        (Top) Pressure anisotropy profiles for (left) an non-grafted film and (right) an chemical identical brush of equal thickness at $\graft = 0.4\,\sigma^{-2}$, for both polymer-liquid-vapor (SL) and polymer-vapor (SV) stacks, at $\eps{pl} = 0.5$.
        The oscillations near $z = 0$ stem from substrate-induced layering.
        (Center) The corresponding interfacial tensions as functions of the lower integration limit $\za$
        in \autoref{eq:interfacial_tension}, for $\zb = 30\,\sigma$.
        (Bottom) The resulting Young's contact angles $\Young$, with the horizontal green line denoting the contact angle $\CA$ measured for droplets.
    }
    \label{fig:sa_detail}
\end{figure*}

\begin{figure}[H]
    \centering
    \includegraphics{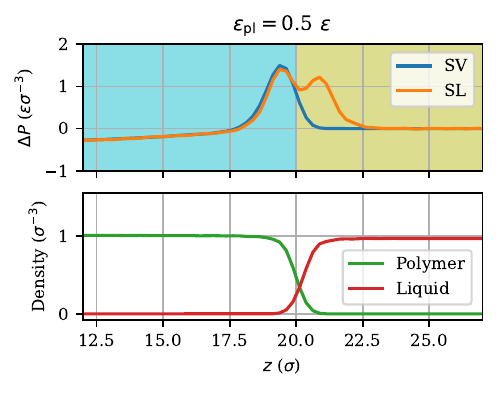}
    \caption{
        (Top) Zoomed in view of the brush-vapor (SV) and brush-liquid (SL) pressure anisotropy profiles of \autoref{fig:sa_detail}, and (bottom) the density profiles of polymeric and liquid particles for the brush-liquid interface.
        The pressure anisotropy profiles $\DP{SV}$ and $\DP{SL}$ almost overlap in the bulk regions, where the polymer concentration is constant, while they differ in the interfacial regions, where the polymer concentration is not constant.
    }
    \label{fig:density}
\end{figure}

\subsection{Interfacial tensions}
The pressure anisotropy of a polymer film and a brush, both with and without a covering fluid layer, are presented in the top row of \autoref{fig:sa_detail}. Since the anisotropy vanishes in the bulk of the liquid layers, the liquid-vapor interfacial tension is readily evaluated as the area under the SL curves, according to \autoref{eq:interfacial_tension}.
The anisotropy also vanishes in the bulk of the polymer film, and consequently the integral over the polymer-vapor (SV) interface and that over the polymer-liquid (SL) interface are both independent of the lower integration boundary $\za$, as long as it lies somewhere in the bulk, as illustrated by wide plateaus in the left-central plot of \autoref{fig:sa_detail}.
In the polymer brush, however, grafting induces a negative pressure anisotropy throughout the bulk that steadily becomes more negative with increasing distance from the surface.
Consequently, the brush-liquid (SL) and brush-vapor (SV) interfacial tensions calculated using \autoref{eq:interfacial_tension} depends on the lower integration boundary $\za$, as shown in the right-central plot of \autoref{fig:sa_detail}.
The plot shows that the values of both $\IntTns{SV}$ and $\IntTns{SL}$ decrease, and even flip sign, when more of the brush bulk is included in the integral, making it impossible to deduce well-defined values for $\IntTns{SV}$ and $\IntTns{SL}$.

\begin{figure*}[bt]
    \centering
    \includegraphics{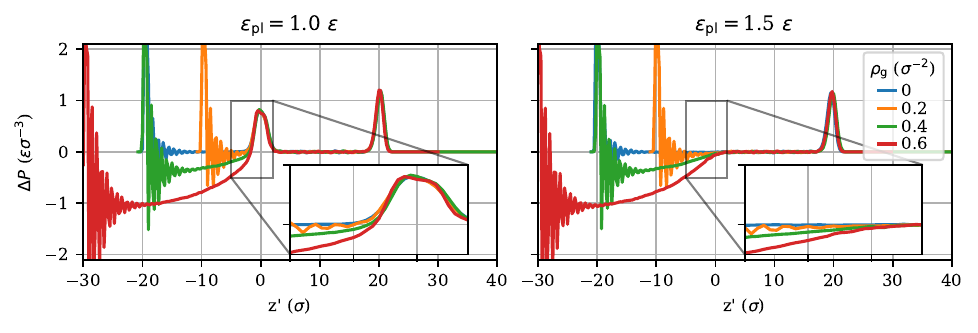}
    \caption{
        Pressure anisotropy profiles for brushes with several grafting densities, and a non-grafted film, $\graft = 0$, in equilibrium with liquid at (left) $\eps{pl} = 1.0\,\epsilon$ and (right) $\eps{pl} = 1.5\,\epsilon$.
        The profiles are presented as functions of the elevation $z' = z - \Hght$ relative to the polymer-liquid interface, with $\Hght$ the height of the polymer
        The oscillations at the left end of the curves, near $z' = - \Hght$, reflect layering the of the polymer particles near the grafting substrate.
    }
    \label{fig:sa_ga}
\end{figure*}

The difference between the two, however, remains remarkably constant, as a consequence of the nearly coalescing $\DP{SV}(z)$ and $\DP{SL}(z)$ curves in the brush. This point is emphasized by the zoomed-in view of the pressure anisotropy profiles in the interfacial region presented in \autoref{fig:density}: the two pressure anisotropy curves are almost indistinguishable at distances of more than $2.5\,\sigma$ from the interface.
Hence, the interfacial tension difference $\IntTns{SV} - \IntTns{SL}$ stems solely from the narrow regions surrounding the brush-vapor and brush-liquid interfaces.
The largest contribution is seen to occur in the liquid, while the pressure anisotropy in the brush is almost identical in the presence and absence of the liquid.
The density distributions in \autoref{fig:density} show that the liquid hardly penetrates the brush; the main difference between the pressure anisotropy profiles occurs in the region where the liquid density is not constant.

Returning to the center row of \autoref{fig:sa_detail}, the constant difference between $\IntTns{SV}(\za)$ and $\IntTns{SL}(\za)$ for the brush over a wide range of $\za$ is nearly identical to the constant difference between their counterparts for the polymer film, because the brush and film share similar interfaces.
We conclude that, although the brush's interfacial tensions $\IntTns{SV}$ and $\IntTns{SL}$ are not well-defined and vary with $\za$, their difference is well-defined and takes on a constant value for an integration boundary $\za$ sufficiently distanced from the interface.

The Young's angles $\theta_y$ obtained by inserting the above surface tension differences in \autoref{eq:IntDPdiff} are plotted in the bottom row of \autoref{fig:sa_detail}.
The weak variations of these angles with the integration boundary $z_a$ are of similar size for brush and film, suggesting that they are probably related to the slow sampling of phase space by polymers.
An excellent agreement is observed with the contact angles $\theta$ measured from the simulations of droplets.
For a collection of films and brushes, the tension differences in the centers of the polymeric layers have been used to calculate the Young's angles plotted in \autoref{fig:CAvsEPS} (dots). The excellent agreement with the contact angles measured from the simulations of droplets validates the use of \autoref{eq:IntDPdiff}, \ie the assumption that $\diff \gamma/\diff \rho_g$ is the same for brush-vapor and brush-liquid.


\subsection{Interfacial and bulk contributions}
We have shown above that the interfacial excess of the pressure anisotropy $\Delta P$ can be identified when considering differences in $\Upsilon_{\rm SV}-\Upsilon_{\rm SL}$. A remaining question, however, is whether we can unambiguously identify the interfacial excess of the pressure anisotropy for a single system, say for a given brush-liquid interface. This would enable a definition of the `true' interfacial tension $\Upsilon^{\rm int}$, separating it from the grafting-induced contribution $\Upsilon^{\rm graft}$. 

\begin{figure*}[bt]
    \centering
    \includegraphics{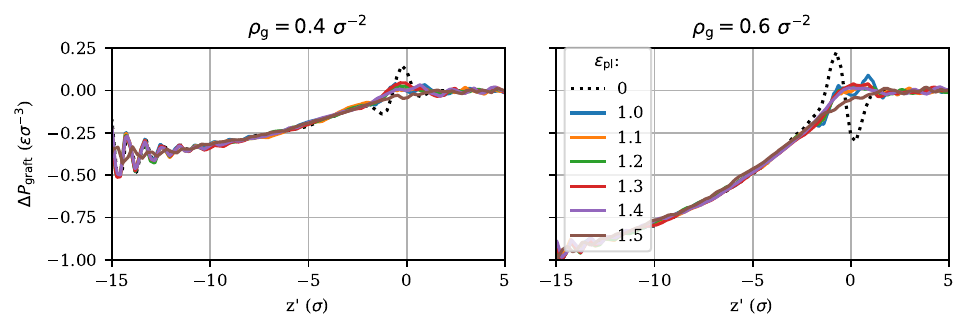}
    \caption{
      Plots of the pressure anisotropy due to grafting, $\DPgraft(z')$,
        for brushes with two grafting densities
            and various polymer-liquid interaction strengths,
          calculated as the difference between the pressure anisotropy of the grafted brush and its ungrafted counterpart at the same $\eps{pl}$.
          Like in \autoref{fig:sa_ga}, the profiles are presented as functions of the elevation $z'$ relative to the polymer-liquid interface. The dashed black line represents the brush-vapor system, which is denoted in the legend by
              $\eps{pl} = 0$.}
    \label{fig:split}
\end{figure*}

To address this question, \autoref{fig:sa_ga} reports the pressure anisotropy profiles $\Delta P(z)$ of several brushes with varying grafting densities for a given polymer-liquid interaction strength. The left panel corresponds to $\epsilon_{\rm pl }= 1.0\,\epsilon$, while the right panel corresponds to $\epsilon_{\rm pl} = 1.5\,\epsilon$, equal to the self interactions. This makes the tensile peak at the interface vanish, resulting in a purely compressive pressure anisotropy profile.
Since the brush thickness varies with the grafting density, the profiles have been shifted with the brush thickness $\Hght$, defined as the inflection point in the polymer density profile, such that their interfacial peaks coincide at $z' = z - \Hght \approx 0$.

The left panel of \autoref{fig:sa_ga} suggests that the total pressure anisotropy of a brush is a sum of two functions of $z'$, according to 
\begin{align}
    \DPbrush( z' )
    \approx
    \Delta P^{\rm graft}_{\graft}(z')
    + \Delta P^{\rm int}_{\eps{pl}}(z').
\end{align}
In this idealization, to be explored in more detail below, the first term on the right hand side is a compressive contribution that significantly varies with the grafting density but is independent of the polymer-liquid interaction, while the second is the tensile interfacial contribution that appears to be similar for all $\rho_g$ at a given $\eps{pl}$.
\SKIP{
It is, however, not possible to extract $\Upsilon^{\rm int}$ as an integral of $\Delta P^{\rm brush}$ over a limited range in $z'$. This becomes clear from the insets of \autoref{fig:sa_ga}: the different curves, corresponding to various grafting densities, do not cross each other at a universal point. Rather, $\Delta P^{\rm graft}_{\graft}(z')$ is non-zero also near the interface.
}
This interfacial contribution is non-zero near the interface only, and hence is readily integrated to obtain the interfacial tension $\Upsilon^{\rm int}$.
The superposition of grafting and interfacial contributions makes it, in general, impossible to extract $\Upsilon^{\rm int}$ as an integral of $\Delta P^{\rm brush}$ over a limited range in $z'$.
Specifically, the height $z_0$ where the anisotropy vanishes, $\Delta P^{\rm brush}( z_0) = 0$, which is sometimes used as an expedient to separate grafting and interfacial contributions, cannot be interpreted as the location where the interface ends.

We proceed by exploring the possibility that the interfacial anisotropy is indeed independent of $\rho_g$. In this construction then, by definition, there is no Shuttleworth effect, \ie $\diff \Upsilon^{\rm int} / \diff \rho_g = 0$.
This allows us to explicitly disentangle $\Delta P^{\rm int}(z')$ and $\Delta P^{\rm graft}(z')$. The former can be determined from the non-grafted film, $\rho_g=0$, and is a function that depends only on $\epsilon_{\rm pl}$. The latter follows by subtraction, $\Delta P^{\rm graft}(z') =\Delta P^{\rm brush}(z') - \Delta P^{\rm int}(z')$. 
The resulting $\Delta P^{\rm graft}(z')$ are shown in \autoref{fig:split}, for two values of $\rho_g$. The profiles closely overlap for a broad range of $\epsilon_{\rm pl}$. In this range we thus find a particularly simple scenario: the compressive part of the pressure anisotropy is solely due to grafting, while the tensile part depends only on the interactions near the interface. We note, however, that this scenario runs into its limits
for the brush in contact with vapor, see the dashed lines marked $\epsilon_{\rm pl}=0$ in \autoref{fig:split}, for which the obtained $\Delta P^{\rm graft}(z')$ shows a more pronounced oscillation near the interface than its counterparts for the brush-liquid systems.
Likewise, a small departure is observed for the limiting case $\epsilon_{\rm pl}=1.5 \,\epsilon$, which is discussed in more detail below. Outside these extreme cases, however, it is possible to disentangle grafting and interfacial contributions by assuming there is no Shuttleworth effect.

\subsection{Autophobic demixing and au\-to\-pho\-bic dewetting}
As shown above, when all bead-bead interaction streng\-ths are equal, \ie $\eps{pp} = \eps{pl} = \eps{ll} = 1.5\,\epsilon,$ the peaks in the pressure anisotropy at the polymer-liquid interface vanish (see the right panel of \autoref{fig:sa_ga}).
In this case an non-grafted film mixes with the liquid, since there is no positive interfacial tension. The polymer brush however, does not mix with the covering liquid layer, even though no tension exists between the two phases. This observation is similar to autophobic demixing, where a liquid layer of polymers does not mix with a brush of identical polymers \cite{Halperin86, Wijmans94, Aubouy93, Maas2002, Pastorino2006, Mensink2019, mensink_role_2021}.
The explanation is that, for this athermal system, stretching the grafted polymer chains to accommodate solvent molecules in the brush loses more conformational entropy than the entropy to be gained by mixing.
It can also be explained in terms of the compressive pressure present in brushes, which prevents any more liquid from entering. Of course, ultimately this argument is equivalent, since the pressure in brushes derives from chain elasticity. 

Due to the surface tension difference $\IntTns{SV} - \IntTns{SL}$ being slightly larger than $\IntTns{LV}$, spreading is favored and the droplet adopts a very shallow shape with pre-wetting films on either side. This system provides an interesting situation: there is a phase separation, but it lacks an interfacial tension between the two phases, \ie no peak of positive $\Delta P$ is observed in the right panel of \autoref{fig:sa_ga}.

\begin{figure}[bt]
    \begin{subfigure}[b]{\linewidth}%
        \caption{}%
        \hfill%
        \includegraphics[width=0.79\linewidth]{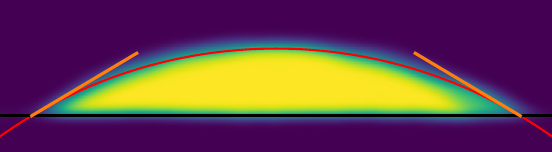}%
        \label{fig:autophobic_dropletfit}%
    \end{subfigure}
    \vspace{1em}
    \begin{subfigure}[b]{\linewidth}%
        \caption{}%
        \includegraphics{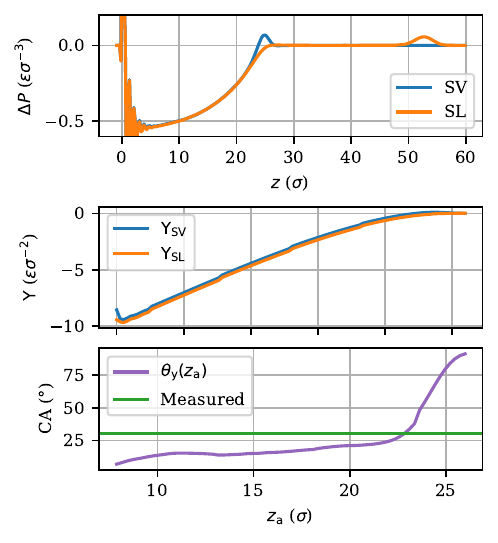}%
        \label{fig:autophobic_SA}%
    \end{subfigure}%
    \caption{
        \textbf{a}: Density profile and circle fit for droplet of 50-mer melt on polymer brush with $\graft = 0.4\,\sigma^{-2}$ and $\eps{pp} = \eps{ll} = \eps{pl} = 0.5\,\epsilon$. \textbf{b}: Corresponding pressure anisotropy profiles (top), interfacial tensions as a function of $\za$ (middle), and resulting Young's contact angle (bottom).
    }
    \label{fig:autophobic}
\end{figure}

Similarly, in autophobic dewetting a polymer melt only partly wets a brush despite a vanishing interfacial tension \cite{Halperin86, Reiter2000, Maas2002, Zhang2008, mensink_role_2021}.
We simulated this phenomenon by using a 50-mer for the wetting liquid, \ie the same chain length as is used for the brush, with all interaction energies reduced to $0.5\,\epsilon$ to relatively enhance the entropic effects.
For this athermal system, a liquid would readily mix when deposited on a non-grafted polymer layer. But the liquid does not mix with a grafted layer, even though the pressure anisotropy does not show a peak at the interface, see \autoref{fig:autophobic}. Still, even though the partial wetting is entropic instead of enthalpic in nature here, due to the surface tension difference $\IntTns{SV} - \IntTns{SL}$ being smaller than $\IntTns{LV}$, the liquid forms a droplet. In other words, the partial wetting is a consequence of $\Upsilon_{\rm LV}$ and $\Upsilon_{\rm SV}$, since $\Upsilon_{\rm SL}$ is never positive.

\section{Conclusion and outlook}
We investigated the influence of grafting-induced negative pressure anisotropy on the wetting of polymer brushes. Focusing on systems where the polymer and liquid do not tend to mix, we systematically compared the interfacial tensions determined by integrating the pressure anisotropy profile of planar polymer-liquid interfaces, and the difference in interfacial energy probed by extracting contact angles from droplets.

The contact angles of droplets on brushes showed no dependence on the grafting density, implying that if this brush system studied here exhibits a Shuttleworth effect, it must be symmetric. Moreover, they agreed closely with the Young's contact angles computed from interfacial tensions. Investigation of the influence of the lower integration limit in the computation of the interfacial tension showed that the negative contribution to the pressure anisotropy induced by grafting appears equally in $\Upsilon_{\rm SL}$ and $\Upsilon_{\rm SV}$. Furthermore, we showed that it is not possible to divide this system in `bulk' and `interface' at a certain point in $z$: while the grafting induced pressure exists throughout the bulk, it also extends into the interface. However, we proposed a route to disentangle the interfacial contribution of the pressure from the grafting-induced pressure, which is applicable over a broad range of parameters.

In this work, we have focused on systems with interaction parameters such that little absorption of the wetting liquid into the brush occurs. Future work might extend this investigation to systems where absorption does occur.


\section*{Author contributions}
\begin{description}
    \item[Lars B. Veldscholte] Conceptualization, software, visualization, investigation, analysis \& interpretation, writing
    \item[Jacco H. Snoeijer] Analysis \& interpretation, writing
    \item[Wouter K. den Otter] Methodology, analysis \& interpretation, writing
    \item[Sissi de Beer] Conceptualization, funding acquisition, resources, analysis \& interpretation, writing
\end{description}

\section*{Declaration of competing interest}
There are no competing interests to declare.

\section*{Acknowledgments}
The authors thank Alvaro González García and Guido Ritsema van Eck for discussions that led to investigating this problem, and Frieder Mugele for his constructive criticism of the manuscript.

This research received funding from the Dutch Research Council (NWO) in the framework of the ENW PPP Fund for the top sectors and from the Ministry of Economic Affairs in the framework of the `PPS-Toeslagregeling' regarding the Soft Advanced Materials consortium. S acknowledges support from NWO Vici (No. 680-47-63).

This work was carried out on the Dutch national e-infrastructure with the support of SURF Cooperative (project EINF-3604).

\section*{Data availability}
The data that support the findings of this study are openly available~\cite{veldscholte_stress_2023}.

\printbibliography

\end{document}